%% file: ICICpaper.tex
\documentclass[runningheads]{llncs}
\usepackage[T1]{fontenc}
\usepackage{graphicx}
\usepackage{booktabs}
\usepackage{threeparttable}
\usepackage{multirow}
\usepackage{amssymb}
\usepackage{amsmath}
\usepackage[colorlinks=true, linkcolor=red, citecolor=blue]{hyperref}
\usepackage{bookmark}
\usepackage{ulem} 
\usepackage{xspace}
\usepackage{tcolorbox}
\usepackage{pifont}
\usepackage{booktabs}
\usepackage{tabularx}    
\usepackage{booktabs}    
\usepackage{multirow}    
\usepackage{array}       
\usepackage{amsmath}
\usepackage{amssymb} 
\usepackage{colortbl}
\usepackage{makecell}
\usepackage{enumerate}
\usepackage{enumitem}
\usepackage{subcaption}
\newcommand{\methodname}{EP-Shield\xspace}
\newcommand{\method}{EP-Shield\xspace}
\newcommand{\methodmetric}{EP-Metric\xspace}
\newcommand{\methodpurifier}{EP-Purifier\xspace}

\begin{document}
\begin{sloppypar}

\title{Evaluate-and-Purify: Fortifying Code Language Models Against Adversarial Attacks Using LLM-as-a-Judge}
\titlerunning{EP-Shield}
%

\author{Wenhan Mu\inst{1}\orcidID{0009-0006-8164-8248}  \and 
Ling Xu\inst{1}\thanks{Corresponding author}\orcidID{0000-0002-7203-511X} \\ \and Shuren Pei\inst{1}\orcidID{0009-0001-3078-1683} \and Le Mi\inst{1}\orcidID{0009-0007-5439-3697} \\ \and Huichi Zhou\inst{2}\orcidID{0000-0002-1518-504X} }

\authorrunning{W. Mu et al.}
%
\institute{\
\inst{1}School of Big Data and Software Engineering, Chongqing University, Chongqing, China  \\
\inst{2}Imperial College London, Exhibition Road, London SW7 2AZ, The United Kingdom \\
\email{whmu@stu.cqu.edu.cn, xuling@cqu.edu.cn, \{pei,mile\}@stu.cqu.edu.cn, \\h.zhou24@imperial.ac.uk }}
\maketitle              
\begin{abstract}
The widespread adoption of code language models in software engineering tasks has exposed vulnerabilities to adversarial attacks,
especially the identifier substitution attacks. Although existing identifier substitution attackers demonstrate high success rates, they often produce adversarial examples with unnatural code patterns.
In this paper, we systematically assess the quality of adversarial examples using LLM-as-a-Judge. Our analysis reveals that 
over 80\% of adversarial examples generated by {state-of-the-art} identifier substitution attackers (e.g., ALERT) are actually detectable. 
Based on this insight, we propose {\method}, a unified framework for evaluating and purifying identifier substitution attacks via naturalness-aware reasoning. 
Specifically, we first evaluate the naturalness of code and identify the perturbed adversarial code, then purify it so that the victim model can restore correct prediction. Extensive experiments demonstrate the superiority of {\method} over adversarial fine-tuning (up to $83.36\%$ improvement) and its lightweight design ($7$B parameters) with GPT-$4$-level performance.


\end{abstract}
\section{Introduction}
\begin{figure}
    \centering
    \includegraphics[width=.7\linewidth]{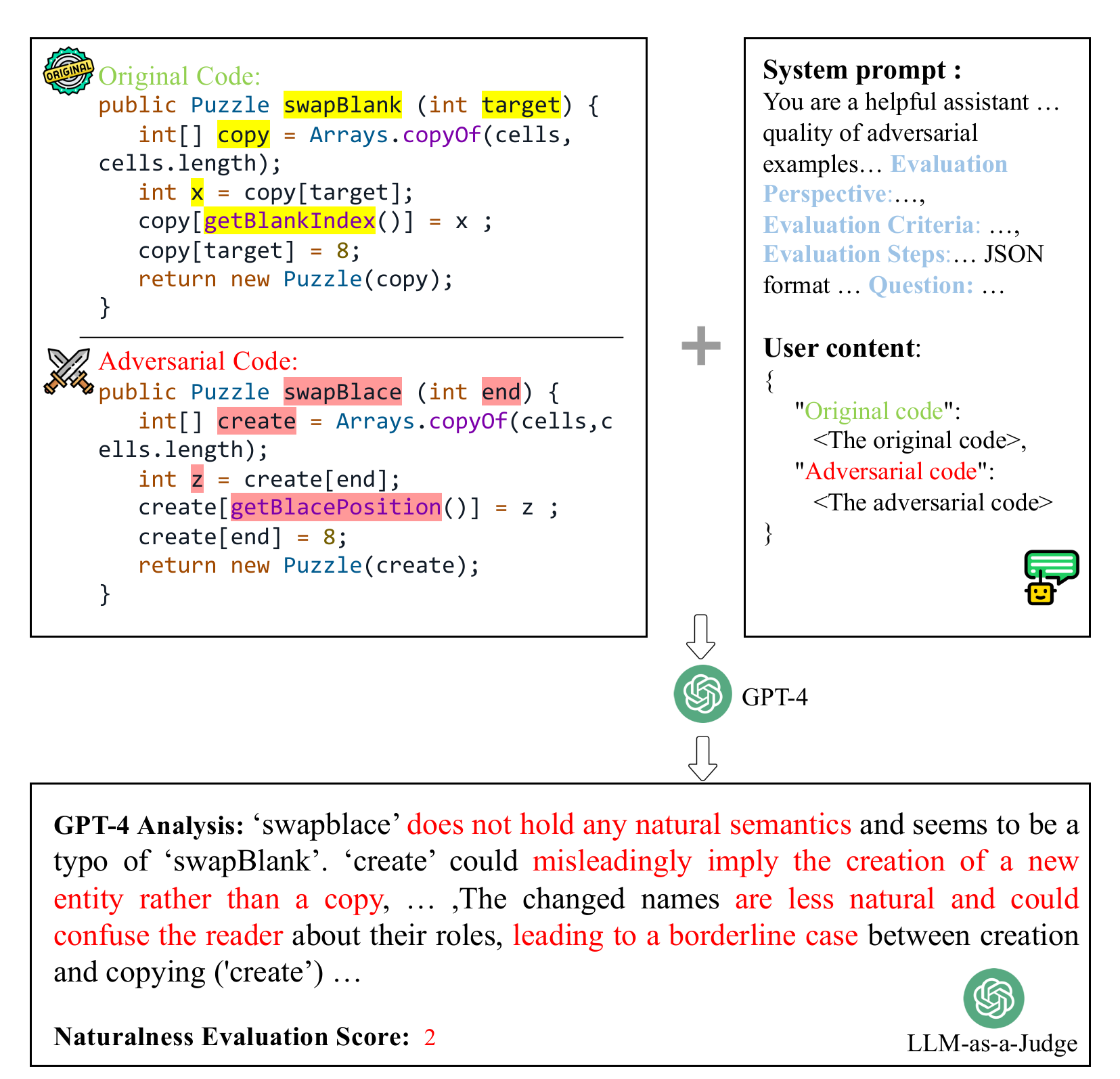}
    \caption{
    A motivating example of an identifier substitution attack, accompanied by an analysis of GPT-4 from the perspective of naturalness using LLM-as-a-Judge.
    }
    \label{fig:motivation-case}
    \vspace{-1em}
\end{figure}

Code Language Models (CLMs), including CodeBERT~\cite{codebert}, PLBART~\cite{plbart}, CodeGPT~\cite{codegpt}, StarCoder~\cite{starcoder}, and Code Llama~\cite{codellama}, have demonstrated remarkable capabilities across a range of software engineering tasks, such as code generation~\cite{code-generation}, code summarization~\cite{code-summarization}, and vulnerability detection~\cite{vulnerability-1}. 
However, CLMs have been shown to be vulnerable to adversarial attacks, such as identifier substitution attacks, where variable or method names are manipulated to mislead models while preserving functionality~\cite{accent,mhm,damp,alert}. 

Most identifier substitution attackers focus on ensuring syntactic validity while overlooking the naturalness of the code~\cite{Regularity-in-Source-Code,code-naturalness,naturalness-of-software,alert}. As shown in Figure~\ref{fig:motivation-case}, the attacker fails to preserve naturalness in two key ways: (1) The logically meaningful identifier \texttt{swapBlank} is mutated to \texttt{swapBlace}, introducing a semantically meaningless subword (\texttt{Blace}); (2) The method name \texttt{copy} is replaced with \texttt{create}, creating a conflict with its associated \texttt{copyof()} function and introducing semantic ambiguity. 
By leveraging the code-understanding capabilities of Large Language Models (LLMs) such as GPT-4~\cite{openai2024gpt4technicalreport}, we evaluate this case using a structured prompt (see Figure~\ref{fig:Prompt for evaluation}). GPT-4 assigns a score of 2 out of 5, reflecting a relatively low level of naturalness and suggesting that the perturbations violate programming conventions.

In this paper, we begin by conducting an empirical study to assess the naturalness of adversarial code generated by state-of-the-art attackers, including MHM~\cite{mhm}, Beam-Attack~\cite{beam}, and ALERT~\cite{alert}.
The experimental results demonstrate that all the current attackers produce adversarial examples with naturalness scores no higher than 2 out of 5, highlighting their unnatural code patterns. This finding emphasizes the critical need to maintain code naturalness when crafting adversarial attacks against CLMs.

Inspired by the success of LLM-as-a-Judge in evaluating code-related tasks~\cite{codeultrafeedbackllmasajudgedatasetaligning,codejudge} and the self-correction ability of LLM~\cite{kumar2024training,zhou2025trustrag}, this paper further puts forward {\method}, a two-stage \textit{evaluate-and-purify} framework to fortify CLMs against adversarial attacks using LLM-as-a-Judge.
We begin by constructing a dataset of $50K$ original-adversarial example pairs accompanied by GPT-4-generated evaluation annotations and a dataset of $46K$ code pairs of adversarial examples and GPT-4 purified counterparts. Based on both datasets, we then fine-tune an LLM (i.e., Qwen) and name it as \method. Specifically, we first evaluate the natural state of a code snippet to determine whether it has been attacked, and purify adversarial examples by transforming them into more natural code.
Finally, the \method can also serve as an evaluation metric to identify adversarial code, and as a purifier to convert the code to benign code. 

To evaluate the effectiveness of \method, we perform experiments over three victim models (i.e., CodeBERT, CodeGPT, and PLBART) across three code-related tasks (i.e., clone detection, code summarization, and vulnerability detection), and compare it against four attack baselines (i.e., MHM, WIR, ALERT, and Beam-Attack).
Experimental results show that \method can achieve defense success rates of approximately 99\%, 95\%, and 90\% in clone detection, vulnerability detection, and code summarization tasks, respectively.
Furthermore, {\method} demonstrates strong generalization ability with over $85\%$ defense success rate against unseen attack patterns. 
Additionally, {\method} achieves strong human evaluation alignment, significantly outperforming baseline metrics.

The contributions of this paper are as follows.
\begin{itemize}[leftmargin=*, itemsep=0pt] 
    \item Using LLM-as-a-Judge, we perform an empirical study to reveal existing identifier substitution attacks to CLMs have the naturalness issue.
    \item 
    We fine-tune a lightweight $7$B-parameter LLM to evaluate and purify the generated adversarial code examples. 
    \item 
    We introduce a naturalness-aware defense strategy that aligns with human code review practices, enhancing the robustness of CLMs against identifier substitution attacks.
\end{itemize}
\section{Background}
\paragraph{Adversarial Attack on CLMs.}
Given a code snippet $x \in \mathcal{X}$, an adversarial attack generates a perturbed sample $x_{adv}$ by applying a slight perturbation $\delta(x)$ such that $\delta: X \rightarrow X_{adv}$. For understanding tasks, a well-trained classifier $f:\mathcal{X} \rightarrow\mathcal{Y}$ should correctly predict the label $y_{truth} \in \mathcal{Y}$, i.e., $f(x) = y_{truth}$, while the adversarial attack misleads the model into producing an incorrect output, i.e., $f(x_{adv}) \ne f(x) = y_{truth}$. For generation tasks like code summarization~\cite{code-summarization}, adversarial examples degrade model performance such that $P(y_{truth}|x_{adv})$ is significantly lower than $P(y_{truth}|x)$. An adversarial example should adhere to the dual channel view of code~\cite{two-channel,codeattack}: \textbf{(1) the formal channel}, ensuring the perturbed code remains syntactically correct and semantically equivalent, and \textbf{(2) the natural channel}, requiring adherence to coding conventions with reasonable identifier names. 

Adversarial code attacks are categorized to three classes: \textbf{(1) Token-level} (e.g., identifier substitution~\cite{alert,accent,beam}), \textbf{(2) Statement-level} (e.g., dead code insertion~\cite{dip}), and \textbf{(3) Structure-level} (e.g., loop/conditional transformations~\cite{coda,ist21,RoPGen}. Token-level attacks dominate due to their high success rate and low detectability~\cite{rnns,beam}, with strategies ranging from random substitution~\cite{wir,mhm} to context-aware methods~\cite{alert,beam}.
\paragraph{Code Naturalness.}
Code exhibits a degree of naturalness analogous to human language, characterized by significant repetitiveness and systematic patterns that can be computationally modeled. This phenomenon arises not only from syntactic constraints inherent to programming languages but also emerges through project-specific conventions, including naming practices and API utilization patterns~\cite{naturalness-of-software,mutants-naturalness,dependency}. Empirical studies~\cite{radar} have demonstrated developers' systematic preference for code variants demonstrating higher predictability. Code naturalness has been attributed not only to syntactic formalism but more fundamentally reflects cognitive predispositions that prioritize ease of code generation and comprehension during software development processes.
\paragraph{Evaluation of Adversarial Examples.}
Prior work focused on attack success rates and efficiency, with ALERT~\cite{alert} introducing naturalness requirements to assess adversarial code, though manual evaluation may introduce bias. Beam-Attack\cite{beam} combines previous metrics~\cite{accent,tkde-naturalness,alert} such as Identifier Change Rate and Average Code Similarity to evaluate attack quality. Traditional tools like SonarQube~\cite{sonarqube} and FindBugs~\cite{findbugs} rely on rule-based identifier analysis, while AST-based methods focus on logic and dependencies, both struggling to detect renaming attacks. Statistical metrics and semantic similarity measures may fail to capture human programming conventions. Increasingly, LLMs are leveraged as evaluation tools~\cite{G-eval,zhou-etal-2024-evaluating-validity,zhang2024can,code-judge-eval,codejudge}.

\section{A Preliminary Study} \label{sec:motivating-study}
\begin{figure*}[t]
  \centering
  \begin{subfigure}{0.32\textwidth}
    \includegraphics[width=\textwidth]{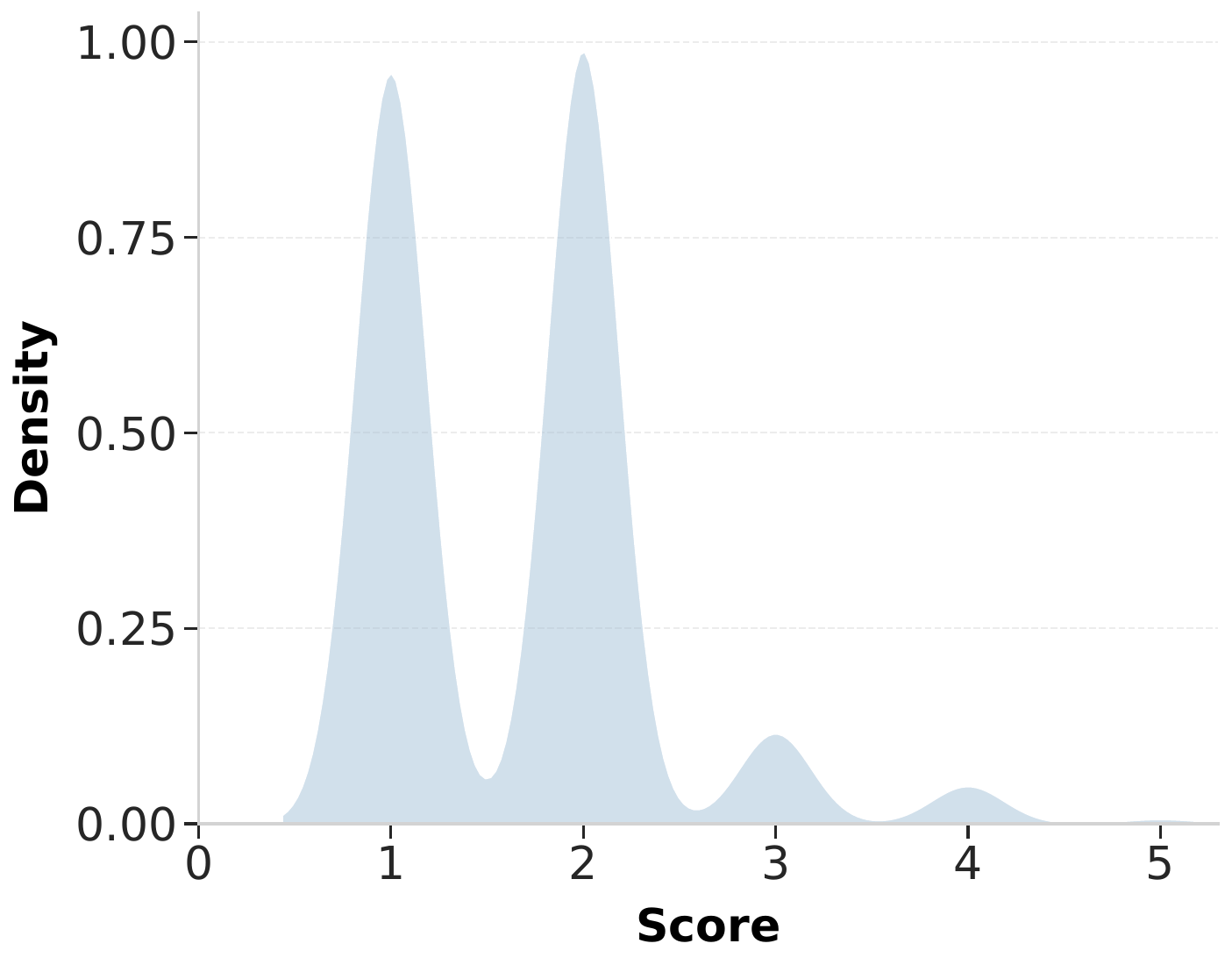}
    \caption{Clone Detection}
    \label{fig:density_sub1}
  \end{subfigure}
  \hfill 
  \begin{subfigure}{0.32\textwidth}
    \includegraphics[width=\textwidth]{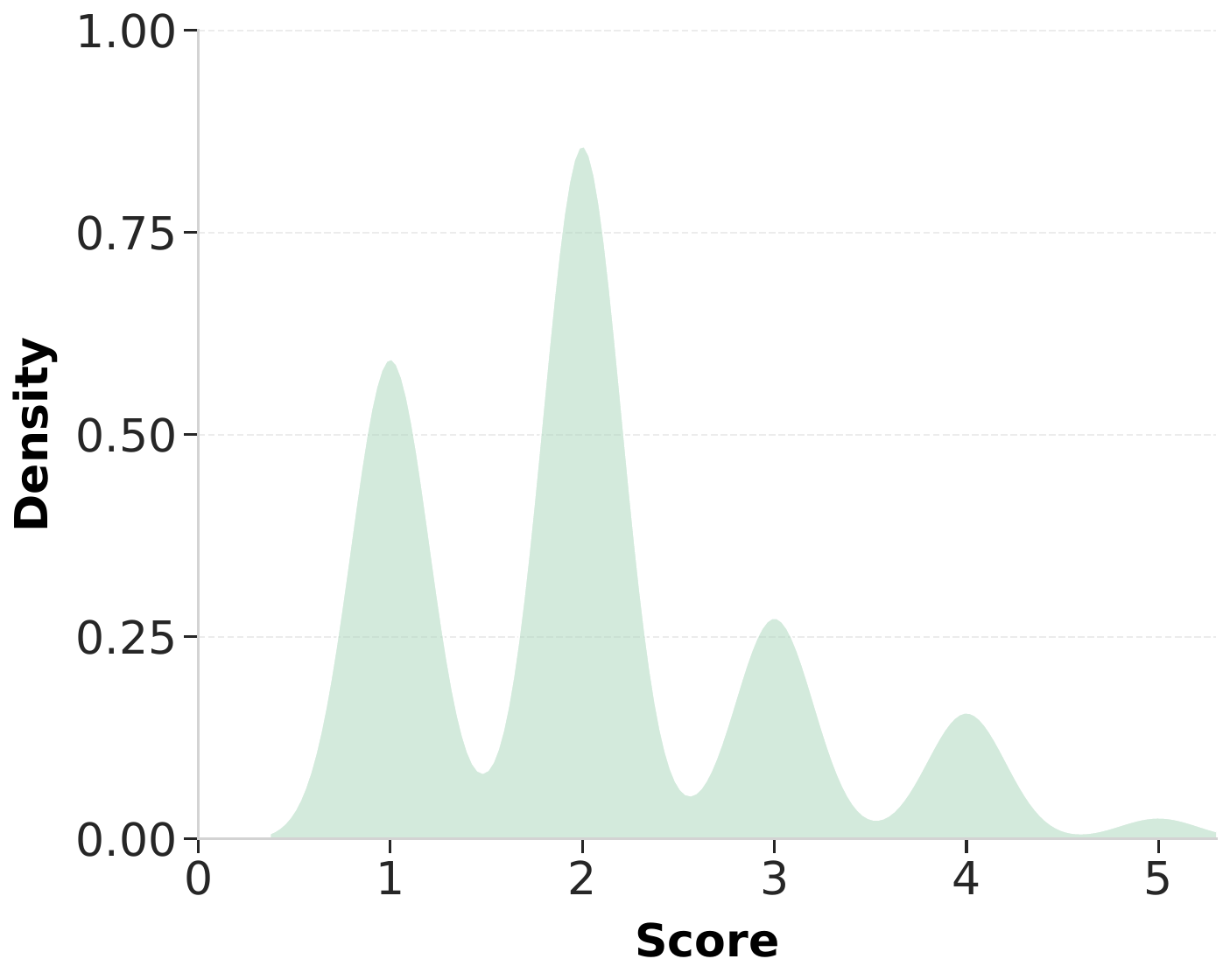}
    \caption{Code Summarization}
    \label{fig:density_sub2}
  \end{subfigure}
  \hfill
  \begin{subfigure}{0.32\textwidth}
    \includegraphics[width=\textwidth]{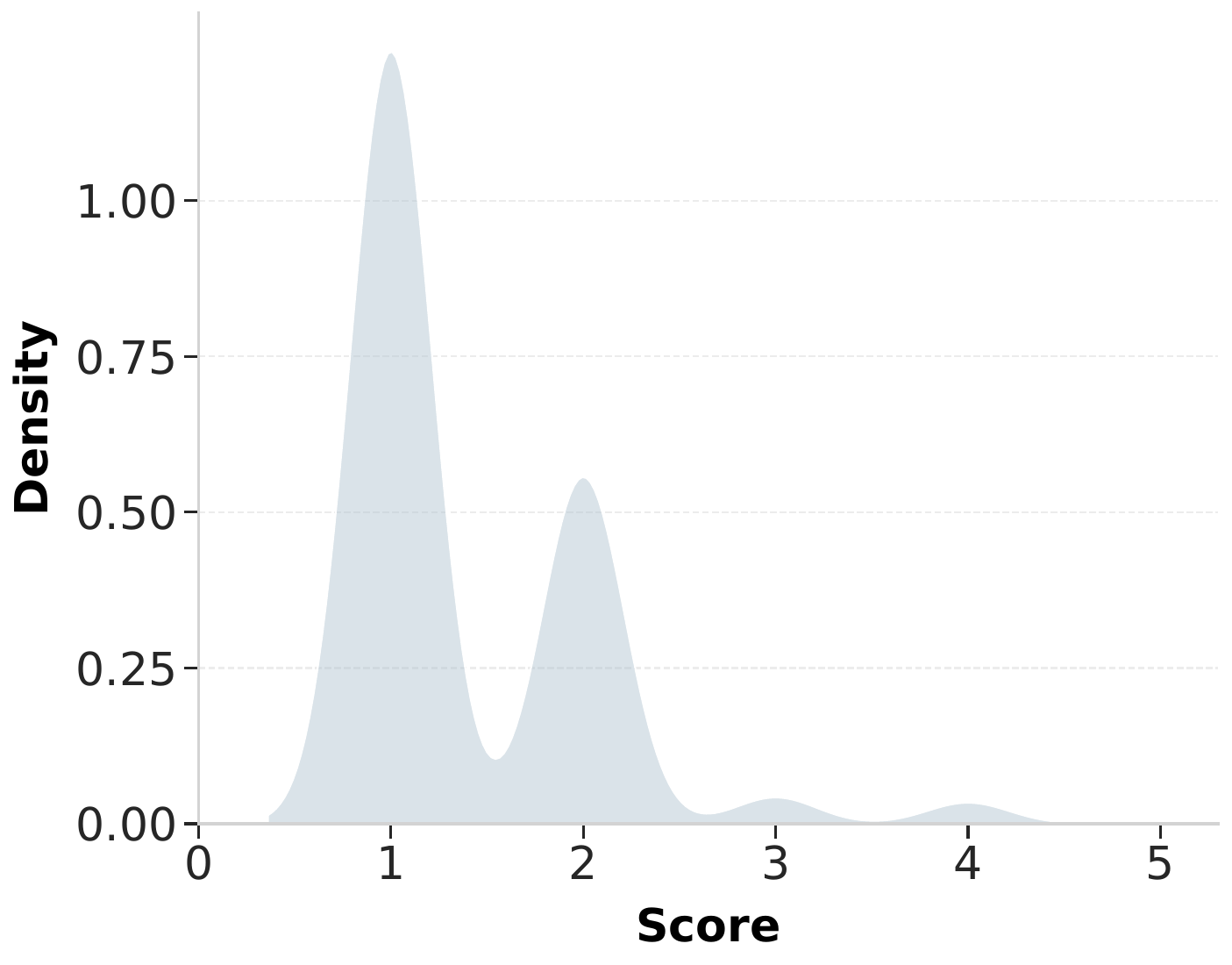}
    \caption{Vulnerability Detection}
    \label{fig:density_sub3}
  \end{subfigure}
  \caption{The density distribution of NES annotated by GPT-4 for adversarial examples generated by ALERT on CodeBERT across three code-related tasks. 
  }
  \vspace{-1em} 
  \label{fig:density_subplots}
\end{figure*}
While existing attack methods claim high success rates, we argue that their generated adversarial examples still exhibit naturalness deficiencies that fail to withstand human scrutiny. 
To validate this hypothesis, we first employ ALERT~\cite{alert} – a state-of-the-art attack method incorporating context-aware naturalness constraints – to attack CodeBERT across three code-related tasks (i.e., clone detection~\cite{clone-detection}, code summarization~\cite{code-summarization}, and vulnerability detection~\cite{vulnerability-1}).

\input{tab/evaluation_score}

We use GPT-4\footnote{We used GPT-4-1106-preview with a temperature of 0 and a top\_$p$ sampling value of 0.9.} to evaluate the naturalness of these adversarial examples, providing analytical annotations with scores ranging from $1$ to $5$, referred to as the Naturalness Evaluation Score (NES). Figure~\ref{fig:density_subplots} presents the density distribution of NES for adversarial examples generated by ALERT across the three aforementioned code-related tasks. As shown in the figure, the probability density at score 1 surpasses 1.00, revealing a highly concentrated distribution pattern. The lower NES indicates that the generated adversarial code is unnatural, which may further suggest an inherent weakness in the attack methodology. This, in turn, could make the adversarial code more easily identifiable.

Using the same way, we investigate the NSE distribution of adversarial examples generated by four attack methods (i.e., ALERT~\cite{alert}, Beam-Attack~\cite{beam}, MHM~\cite{mhm}, WIR~\cite{wir}) across three victim models (i.e., CodeBERT~\cite{codebert}, CodeGPT~\cite{codegpt}, PLBART~\cite{plbart}) and the aforementioned three downstream task combinations. Table~\ref{tab:eval_score} presents the distribution proportions of NES scores ($1$–$5$) and the weighted average scores for each attack scenario. It can be observed that the vast majority of samples exhibit NES scores in a lower range ($\le2$). This suggests that existing attack methods frequently produce unnatural adversarial examples, which may be a common phenomenon.

\section{\method: Our Approach}

\begin{figure*}[!t]
    \centering
    \includegraphics[width=1\linewidth]{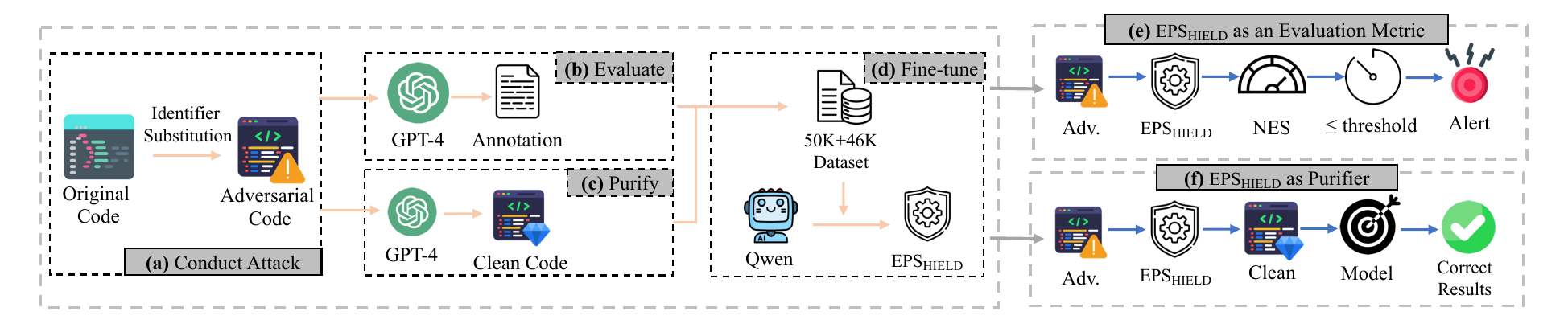}
    \caption{An overview of our proposed {\method}. 
    }
    \vspace{-1em}
    \label{fig:Framework-of-EPIC}
\end{figure*}

Figure~\ref{fig:Framework-of-EPIC} shows an overview of our proposed {\method}.
We first conduct identifier substitution to collect adversarial code (Figure~\ref{fig:Framework-of-EPIC}(a)).
Subsequently, we conduct a naturalness analysis of adversarial code with the assistance of LLMs and collect annotations (Figure~\ref{fig:Framework-of-EPIC}(b)). 
In addition, we employ an LLM to purify adversarial code and collect the resulting benign code after this process (see Figure~\ref{fig:Framework-of-EPIC}(c)).
Furthermore, we use the dataset collected from the aforementioned two steps to fine-tune a smaller LLM, resulting in \method (Figure~\ref{fig:Framework-of-EPIC}(d)).

The obtained \method can serve as an evaluation metric to identify adversarial code (Figure~\ref{fig:Framework-of-EPIC}(e)), We refer to this role as \methodmetric.
Additionally, \method can also serve as a purifier to convert adversarial code into benign code to restore the correct predictions of the target model  (Figure~\ref{fig:Framework-of-EPIC}(f)), We refer to this role as \methodmetric.

\subsection{Identifier Substitution Attack}
To systematically investigate identifier substitution attacks, we generated a dataset of over $50K$ adversarial examples by applying four different attack methods across three datasets and three victim models. These adversarial examples serve as the foundation for subsequent evaluations in \S~\ref{subsec:method-evaluation} and~\ref{subsec:method-purification}. To ensure a comprehensive analysis of the attack characteristics, we retained only successful adversarial examples where the target model produced incorrect outputs due to identifier substitutions while maintaining functional equivalence. This dataset enables us to capture variations in attack effectiveness and adversarial code patterns.

\subsection{Naturalness Evaluation via LLMs}\label{subsec:method-evaluation}
Based on the filtered adversarial examples, we query GPT-4 in an open-ended manner, asking it to evaluate each adversarial example. We instruct GPT-4 to evaluate based on the following analytical perspectives: 
    \begin{tcolorbox}[
      width=\linewidth,
      colback=gray!00,
      colframe=black,
      arc=1.5mm,
      auto outer arc,
      left=2mm,
      right=2mm,
      boxrule=0.9pt,
      colbacktitle=black!65!black
    ]
    \small
    \textit{(1) From the perspective of the natural semantics of identifiers, how correlated is each identifier's name with its role in the program?}\\
    \textit{(2) In terms of readability and naturalness, does each identifier's name seem reasonable and as if chosen by an experienced programmer in a normal coding scenario?}
    \end{tcolorbox}

Figure~\ref{fig:Prompt for evaluation} shows the design details of the prompt templates for this task. We prompt GPT-$4$ to perform analyses in a Chain-of-Thought~\cite{chain-of-thought} (CoT) manner to achieve a deep understanding of naturalness consistency, thereby making the evaluation scores interpretable. To provide directions for GPT-$4$'s reasoning, we designed the steps necessary in the prompt template. The ``System prompt'' includes the reasoning perspectives, processing steps, some output constraints, and specific task details. The ``User content'' provides the original adversarial example pair to be evaluated. 

For the evaluation task, we require GPT-4 to annotate each adversarial example with a score from $1$ to $5$ and provide a corresponding analysis. This score is the Naturalness Evaluation Score(NES) defined above. The ``Analysis'' serves to justify the assigned score by outlining GPT-4’s reasoning process, ensuring interpretability and transparency. Additionally, this explanation provides a foundation for fine-tuning the local LLM. Ultimately, GPT-4’s output will be structured as follows:
\begin{tcolorbox}[
      width=\linewidth,
      colback=gray!00,
      colframe=black,
      arc=1.5mm,
      auto outer arc,
      left=2mm,
      right=2mm,
      boxrule=0.9pt,
      colbacktitle=black!65!black
    ]
    \small
    \textbf{Analysis:} \textit{<Analysis of the annotated score.>} \\    
    \textbf{Score:} \textit{<Predicted integer score (1 to 5).>} \\    
    A score of 1 indicates significant unnatural issues in the code segment, while a score of 5 reflects high adherence to common programming practices, ensuring readability and naturalness.
    \end{tcolorbox}

Finally, we remove a small number of examples exceeding the maximum input length limit ($4096$ tokens) of the local LLM, resulting in an instruction dataset containing over $50K$ examples for evaluating the naturalness of adversarial examples.
The ``System prompt'' and ``User content'' in the prompt template correspond to the instruction and input in the instruction dataset, respectively, while GPT-$4$'s output corresponds to the dataset output. This dataset can facilitate research on evaluating the effectiveness of identifier substitution attacks. 

\subsection{Code Purification via LLMs}\label{subsec:method-purification}
We also employ GPT-4 to refine the collected adversarial code from a naturalness perspective, thereby preventing it from misleading the victim models.
Figure~\ref{fig:Prompt for purfication} shows the design details of prompt for this purification task. Here, GPT-4 adopts the CoT reasoning approach and aligns with the same analytical perspectives as those used in the evaluation task. In this task, the output consists solely of the purified code. To ensure effectiveness, all purified code must be re-evaluated by the victim model, filtering out unsuccessful purification attempts. This step prevents the introduction of noise during the fine-tuning stage, maintaining the integrity of the training data. 
Ultimately, we obtain an instruction dataset containing over $46K$ samples for further research on the purification of adversarial code. The structure of this dataset is consistent with the dataset acquired during the evaluation task.

\subsection{LLM Fine-Tuning}
Instruction fine-tuning based on GPT-$4$ is a popular and effective fine-tuning approach~\cite{xu2024surveyknowledgedistillationlarge}. Benefiting from the recent emergence of numerous open-source LLMs~\cite{li2023textbooksneediiphi15,codellama}, lighter-weight alternatives have emerged, making it feasible to replace GPT-$4$. Meanwhile, the previously mentioned instruction dataset equips local LLMs with essential background knowledge and references. Through fine-tuning, it further mitigates the challenges small LLMs encounter when handling complex tasks.

In this work, we choose Qwen$2.5$-coder-$7$B~\cite{qwen2.5coder}, a lightweight and open-source model, as the backbone of {\method}. We fine-tune it using the instruction-response data obtained from the previous two sections.

Our goal for {\method} is to generate reasonable explanations $e$ for the NES of adversarial examples during the evaluation task. The optimization objectives for supervised fine-tuning are as follows:
\begin{equation}
    \mathcal{L}_{e}=\mathbb{E} \log \mathrm{P}_{\mathrm{{\method}}}\left([e, \mathrm{NES}] \mid\left[\texttt{Ins}_{e}, c_{o}, c_{a}\right]\right)
    \label{equation:optimization-obj-eval}
\end{equation}
where $c_{o}$, $c_{a}$, and $c_{p}$ represent the original code, perturbed code, and purified code, respectively.
For the purification task, we expect {\method} to identify locations with unnaturalness and generate appropriate purified code. The optimization objectives are as follows:
\begin{equation}
    \mathcal{L}_{p}=\mathbb{E} \log \mathrm{P}_{\mathrm{{\method}}}\left( c_{p} \mid\left[\texttt{Ins}_{p}, c_{o}, c_{a}\right]\right)
    \label{equation:optimization-obj-puri}
\end{equation}
where $\texttt {Ins}_{e}$ and $\text {Ins}_{p}$ are identical to the instructions given to GPT-4 for the two tasks.

\subsection{{\method} as an Evaluation Metric}
As shown in Figure~\ref{fig:motivation-case}, the example obtained after existing identifier substitution attacks exhibit a certain degree of unnaturalness. The fine-tuned {\method} will learn the characteristics of unnatural perturbations and provide an interpretable evaluation score. This human-aligned evaluation score can be a new constraint for identifier substitution attacks. 
As shown in Figure~\ref{fig:Framework-of-EPIC} (c), when the NES of adversarial code falls below a predetermined threshold, it serves as an indication that the input code is likely to have been attacked.
It is important to note that in real-world scenarios, we only receive a piece of code as input, and must rely on the scores and explanations provided by {\method} to determine whether it is adversarial code. Therefore, during the inference stage, the prompt template should be adjusted accordingly for the input description.

\subsection{{\method} as a Purifier}
The potential naturalness flaws in adversarial code resulting from identifier substitution attacks can also serve as a breakthrough for defending against similar attacks. The fine-tuned {\method} will follow the analysis steps outlined in the instruction template to locate identifiers with unnaturalness, and then replace them based on their functionality and the overall code semantics, with names that align more closely with human programming conventions, allowing CLMs to recover their original correct prediction results.

\begin{figure*}[t]
    \centering
    \begin{tcolorbox}[
                  colback=gray!00,
                  colframe=black,
                  width=\textwidth,
                  arc=1.5mm, auto outer arc,
                  left=0.9mm, right=0.9mm,
                  boxrule=0.9pt, colbacktitle = black!65!black]
    \textbf{System prompt:}\\ You are a helpful assistant in analyzing and measuring the quality of adversarial examples: you will be given an original code snippet and an adversarial code snippet with only the identifier name replaced. You need to follow the following evaluation perspective and evaluation criteria to rate each adversarial sample.
    
    \ding{228} \underline{\textit{Evaluation Perspective}:}\\
    1. At the level of the identifier's natural semantics, how relevant is the replaced identifier's name to its role in the program? The higher the degree of such relevance, the higher the score should be given on this dimension. \\
    2. In terms of readability and naturalness, check whether the changed identifier name looks reasonable from the perspective of an experienced programmer and whether it looks like an identifier name that may be used in a normal programming environment. Similarly, the stronger the naturalness, the higher the score of this aspect should be.\\
    \ding{228} \underline{\textit{Evaluation Criteria}:}\\
    Use 1-5 points to measure the comprehensive score of the adversarial sample on the two contents mentioned in the "Evaluation Perspective." A score of 5 means that the identifier name after replacement maintains the consistency of natural semantics and actual functions and the naturalness of the name. A score of 1 indicates that the consistency and naturalness of the identifier name after replacement are very low.
    
    \ding{228} \underline{\textit{Evaluation Steps}:}\\
    1. Read both the original code and adversarial code examples carefully. \\
    2. Pay attention to the identifiers after replacement in the adversarial examples. I will give you the code snippets you need to analyze in the following JSON format:\\
    \{ \\
    "Original code": <The original code>, \\
    "Adversarial code": <The adversarial code>\\
    \} \\
    3. When answering, please keep the analysis brief but make sure the explanation is clear. Meanwhile, give a numerical score.\\
    \ding{228} \underline{\textit{Question}:} 
    \\On a scale of 1-5, where 5 is the highest, How well do adversarial examples satisfy the two evaluation perspectives? You should follow the Evaluation Steps and rate based on the Evaluation Criteria. Please strictly follow the following JSON format for output:\\
    \{\\
    "Analysis": <Your analysis>,\\
    "Score": <Your score for adversarial example> \\
    \}\\
    \textbf{User content:} \\
    \{ "Original code": <The original code>, "Adversarial code": <The adversarial code> \} 
    \end{tcolorbox}
    \caption{The prompt template for LLMs to obtain the NES and analysis.}
    \label{fig:Prompt for evaluation}
\end{figure*}

\begin{figure*}
    \centering
    \begin{tcolorbox}[colback=gray!00,
                  colframe=black,
                  width=\textwidth,
                  arc=1.5mm, auto outer arc,
                  left=0.9mm, right=0.9mm,
                  boxrule=0.9pt, colbacktitle = black!65!black]
    \textbf{System prompt:} \\
    You are a helpful assistant to help code pre-train models against adversarial code attacks: The input to the victim model is a piece of code. By purifying the adversarial code snippets, the purified code snippets will not output incorrect results when input into the model.\\
    \underline{\textit{Purification steps}:}\\
    1. Read the adversarial code snippet carefully.\\
    2. Locate identifiers that you think were replaced (attacked) in the original code.\\
    3. Replace the names of these identifiers with names that you think the original code might have used or that you think are reasonable.\\
    \underline{{\textit{You can refer to the following perspectives for purifying}:}}\\
    1. At the level of the natural semantics of an identifier, how relevant is the name of an identifier to its role in the program? \\
    2. In terms of readability and naturalness, check that the name of the identifier is reasonable from an experienced programmer's point of view and looks like an identifier name that might be used in a normal programming environment.\\
    \underline{\textit{You must strictly abide by the following rules}:}\\
    1. You can only make changes to identifier names, not to other parts of the code.\\
    2. The modified code must be syntactically correct and compilable.\\
    3. Output only your modified code itself and do not include anything else.\\
    \textbf{User content:} \\
    \{ <The adversarial code> \} 
    \end{tcolorbox}
    \vspace{-1em}
    \caption{The prompt template for purification task.}
    \label{fig:Prompt for purfication}
    \vspace{-1em}
\end{figure*}

\section{Experiments and Analysis}
\label{sec:experiment setup}
We study the following research questions:
    \textit{(1) Is {\method} suitable for evaluating adversarial code naturalness?}
    \textit{(2) Is {\method} a better measure of adversarial example naturalness?}
    \textit{(3) How effective is {\method} as a purifier in enhancing model robustness?}
    \textit{(4) Does {\method} mistakenly identify clean examples?} 
    \textit{(5) Is {\method} still effective when facing unknown patterns?}
    We also perform a case study to show the effectiveness of our proposed approach.

\subsection{Experimental Setup}

\input{tab/statisitcs_of_datasets}

\paragraph{Datasets.} We select three representative code processing tasks: \textit{clone detection}, \textit{vulnerability detection}, and \textit{code summarization}. Clone detection identifies functionally similar or identical code despite syntactic differences, using the BigCloneBench~\cite{bigclonebench} dataset, processed by ~\cite{alert}, containing $90,102$ training samples and $4,000$ validation and test samples each. Vulnerability detection discovers security vulnerabilities in software, using the OWASP benchmark, which includes $13,041$ training samples and $4,000$ validation and test samples each. Code summarization generates natural language summaries for code, using a subset of CodeSearchNet~\cite{codesearch}, processed following prior work~\cite{codegpt,wir}, containing $164,923$ training samples, $5,183$ validation samples, and $10,955$ test samples. The dataset statistics corresponding to the tasks are shown in Table~\ref{tab:statisitcs of datasets}. To assess {\method}'s generalization ability across unknown data distributions, we performed authorship attribution experiments on the Google Code Jam dataset.

\paragraph{Victim Models.} We selecte representative CLMs from three categories of structures (encoder-only, decoder-only and encoder-decoder): CodeBERT, CodeGPT, and PLBART. These models were fine-tuned on the three downstream tasks above (referred to as Clone Detection, Vulnerability Detection, and Code Summarization). The reproduced results are shown in Table~\ref{tab:performance of CLMs}.
\input{tab/performance_of_CLMs}

\input{tab/attack_performance}
\paragraph{Attack Methods.} We select four major attack methods using identifier substitution. MHM~\cite{mhm} iteratively renames identifiers using Metropolis-Hastings sampling~\cite{mhsample}. WIR-Random~\cite{wir} selects identifiers based on Word Importance Rank~\cite{textattack} and replaces them randomly. ALERT~\cite{alert} applies contextual naturalness-aware substitutions using a greedy and genetic algorithm. Beam-Attack~\cite{beam} leverages statement type knowledge and beam search~\cite{beamsearch} for identifier selection. Attack success rates for each method are shown in Table~\ref{tab: ASR of methods}.

\paragraph{Metrics.} We use seven baseline metrics to evaluate adversarial code quality and code similarity, with Identifier Change Rate (ICR) and Token Change Rate (TCR) measuring identifier and token modifications, respectively; Average Code Similarity (ACS) and Average Edit Distance (AED) assessing code similarity and token-level differences; Code Perplexity (PPL) quantifying model prediction accuracy; and CodeBLEU~\cite{codebleu} and CodeBERTScore~\cite{codebertscore} evaluating syntax, semantics, and embedding-based similarity.

\input{tab/consistency_analysis}

\subsection{Validity in Naturalness Assessment (RQ1)}

We use half of the instruction fine-tuning dataset collected for evaluation in~\S~\ref{subsec:method-evaluation} as a training set to fine-tune {\methodmetric}, enabling it to perform the same evaluation tasks as GPT-4 and learn how to analyze the naturalness flaws of adversarial examples. We then test it on the remaining half of the dataset. 
We invited six non-author practitioners to manually annotate the same test data to demonstrate the feasibility of our proposed evaluation metrics and the effectiveness of the {\methodname}. These annotators are all computer science professionals with at least three years of programming experience, representing diversity in gender, age, and educational background. 
They are instructed to provide objective judgments without considering the length of the answers, aiming to minimize human bias. We conduct Mann-Whitney U tests~\cite{Mann–Whitney-U} to compare the NES produced by GPT-$4$, {\methodmetric}, and human evaluators at different analytical perspectives. 

Table~\ref{tab:consistency-analysis} shows the results of the consistency tests, We can observe that both GPT-$4$ and {\methodmetric} passed the consistency check with human evaluators across all attack combinations ($p$-value $< 0.05$), demonstrating a high level of consistency with human assessments. Overall, the consistency of {\methodmetric} with human evaluations is very close to that of GPT-$4$, achieving a 65\% exact match rate and a 95\% consistency rate within a tolerance of $1$. This suggests that {\methodmetric} can effectively simulate human judges to evaluate the naturalness of adversarial examples from a human perspective, serving as a new metric for assessing adversarial example quality. 
This also indicates that the scoring results from GPT-$4$ in~\S~\ref{sec:motivating-study} are reasonable, and indirectly confirms that current identifier substitution attacks exhibit naturalness issues.

\input{tab/correlation_overall}

\subsection{Comparison with Existing Metrics (RQ2)}
We compare {\methodmetric} against seven baseline metrics using human judgments as ground truth (Table~\ref{tab:correlation metric vs human}). Our method achieves the highest average correlation (0.6831) across Spearman (0.6687), Pearson (0.7419), and Kendall-Tau (0.6386) coefficients, demonstrating superior alignment with human assessments of code naturalness. This consistent performance underscores {\methodmetric}'s effectiveness in simulating human reasoning through its comprehensive evaluation framework. Notably, CodeBLEU and Code Perplexity (PPL) exhibit negligible correlations (Avg. 0.0414 and analogous trends). 
Semantic-aware metrics (ACS, AED, CodeBERTScore) outperform statistical measures (TCR, ICR), suggesting semantic shifts impact naturalness more than token-level changes. However, {\methodmetric} uniquely combines three critical dimensions: (1) local semantic preservation, (2) global code coherence, and (3) adherence to programming conventions. This multi-perspective approach enables finer-grained detection of unnatural modifications compared to single-aspect baselines.

\input{tab/purify_compare_subset}
\subsection{ Performance of Defense (RQ3)}

\label{sec:defense effectiveness}


We fine-tune {\methodpurifier} using $50\%$ of the purification dataset from~\S~\ref{subsec:method-purification}, evaluating robustness on the remaining data against GPT-$4$ and adversarial tuning (Adv-tuning)~\cite{damp}. Table~\ref{table:defense compare on codebert} presents the defense success rates of {\methodpurifier}, GPT-$4$, and Adv-tuning on CodeBERT (Further details are provided in Table~\ref{table:defense compare}).

Both tables reveal that {\methodpurifier} achieves near-GPT-4 defense success rates ($\le 2.5\%$ gap) while significantly outperforming Adv-tuning. Particularly in code summarization tasks, our method defends $36.19\%$ more adversarial examples than Adv-tuning against MHM attacks on CodeGPT, demonstrating superior defensive capabilities. Despite showing marginal improvement ($0.27\%$) over {\methodpurifier} in vulnerability detection with PLBART, Adv-tuning exhibits critical instability — its defense rate plummets to $20.82\%$ in CodeGPT-based clone detection, contrasting with {\methodpurifier}'s consistent robustness across all tasks. This performance divergence highlights the limitations of adversarial retraining approaches compared to our purification paradigm.

\subsection{ Misclassification Rate (RQ4)}

\input{tab/misclassfication_rate}

We randomly select 1,000 examples from the training set of each of the three datasets and used {\methodmetric} to annotate their NES scores. Table~\ref{tab: misclassfication-rate} presents the proportion of clean examples that {\methodmetric} incorrectly classifies as adversarial under different thresholds $\delta$. 
When $\delta = 2$ (i.e., adversarial examples are those with NES $\le 2$, {\methodmetric} misclassifies only a small number of clean examples. Corresponding to Table~\ref{tab:eval_score}, most adversarial examples have NES values below or equal to $2$, suggesting that setting the threshold to $2$ may be a good criterion for classification. As $\delta$ increases, the misclassification rate also rises, particularly when $\delta = 4$. In this case, the misclassification rates for the three tasks reach $64.30\%$, $60.52\%$, and $62.27\%$, with a significant portion of examples having NES values around $4$. This indicates that a stricter threshold for detecting adversarial examples can lead to a higher possibility of misclassification.

\subsection{Generalization Ability Analysis (RQ5)}

We evaluate {\methodpurifier}'s generalization capability against unseen attack patterns and tasks. First, we test on RNNS~\cite{rnns}—a SOTA identifier-level attack—for vulnerability detection. As shown in Table~\ref{tab:RNNS-VD}, {\methodpurifier} outperforms Adv-tuning across all models, with improvements up to $31.34\%$ (CodeGPT). Notably, {\methodpurifier} maintains $>85\%$ defense success despite RNNS's advanced perturbation strategy.
Second, we extend evaluation to Authorship Attribution~\cite{authorship-attribution2017}, a code style analysis task. Table~\ref{tab:RNNS-AS} shows {\methodpurifier} surpasses Adv-tuning by $8.05-12.33\%$ against three attack methods, even when both dataset and attack mechanism are unseen. 
Both experimental results demonstrate that {\methodpurifier}, stemming from natural perceptual reasoning, maintains defensive capabilities even in unknown scenarios while preserving robustness.

\input{tab/defence4unknown}

\input{tab/defence4unknown_2}

\subsection{Case Study}
\begin{figure}
    \centering
    \includegraphics[width=\textwidth]{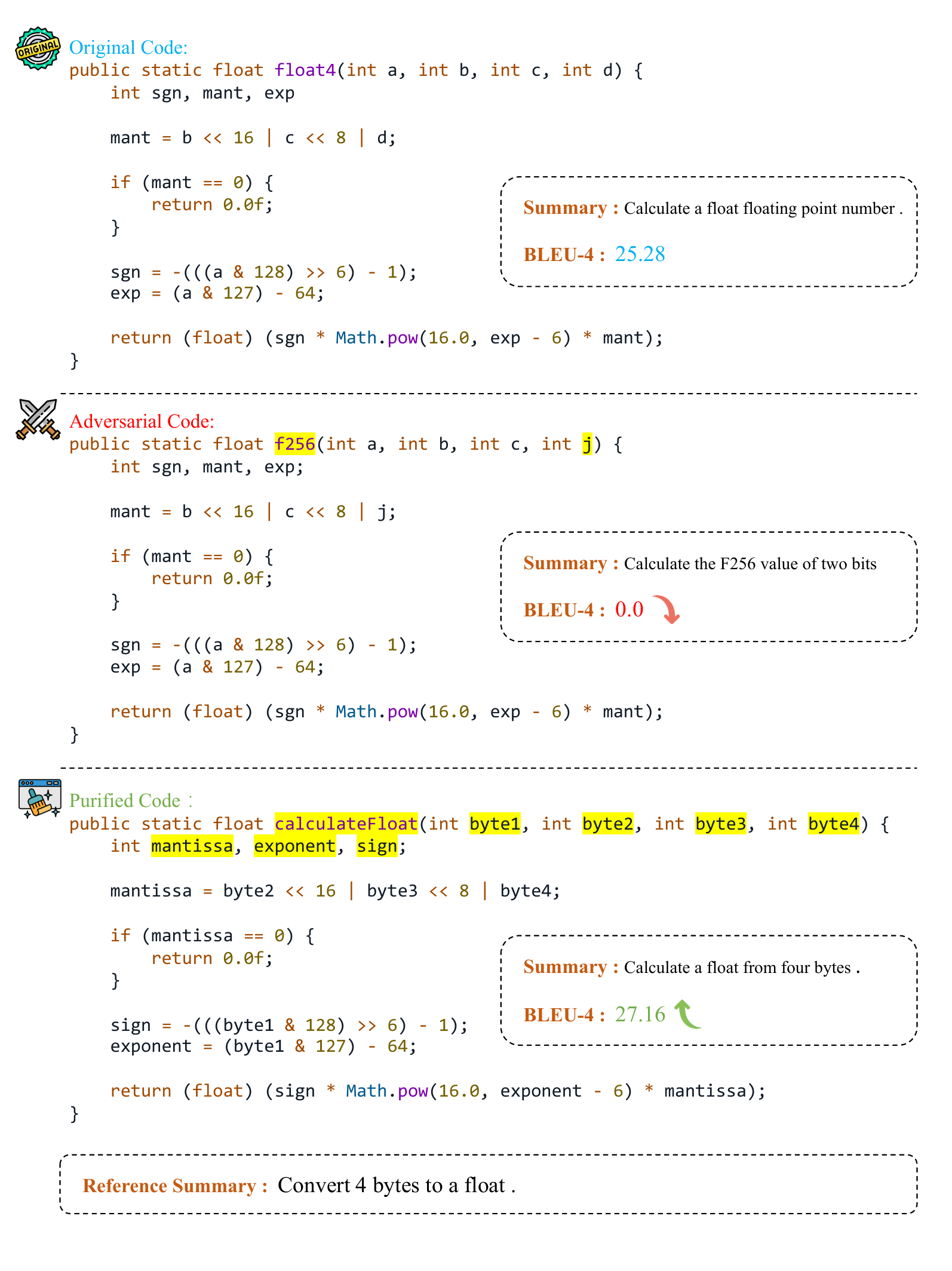}
    \caption{
    Case study on code purification. 
    }
    \label{fig:case-study}
    \vspace{-1em}
\end{figure}

As shown in Figure~\ref{fig:case-study}, which illustrates three stages: original, adversarially perturbed, and \method-purified code. Each snippet performs the same task: converting four bytes into a floating-point number using bitwise operations. The attacker replaces meaningful identifiers (e.g., `\texttt{float4}'' $\rightarrow$ \texttt{F256}'' and `\texttt{d}'' $\rightarrow$ \texttt{j}''). These subtle changes cause semantic confusion, resulting in an incorrect summary (``Calculate the F256 value of two bits'') with a BLEU-4 score of $0.0$, highlighting how identifier substitution degrades model performance while maintaining syntactic validity. In contrast, \methodname{} restores meaningful identifiers (e.g., `\texttt{mant}'' $\rightarrow$ \texttt{mantissa}'') and follows naming conventions (e.g., `\texttt{calculateFloat}''). The purified code improves readability and summarization quality (BLEU-4: $27.16$ vs. $25.28$), generating ``Calculate a float from four bytes,'' which better aligns with the reference summary. This case demonstrates the framework's value in maintaining model robustness and code interpretability.

\input{tab/purify_compare}

\section{Extended Analysis}\label{Extended-Analysis}

\subsection{Pre-Fine-Tuning vs. Post-Fine-Tuning}
\input{tab/preft_and_postft}
We compare the defense success rates of {\methodname} before and after fine-tuning. The results are summarized in Table~\ref{tab: pre-ft-vs.-post-ft}. Results show that {\methodname} significantly enhances model robustness against various adversarial attacks. For example, CodeBERT's success rate under ALERT increased from $80.51\%$ to $98.89\%$, and under WIR from $73.35\%$ to $99.27\%$. Similar improvements are observed for CodeGPT and PLBART, demonstrating that {\methodname} effectively learns to counter adversarial perturbations, enabling victim models to better resist attacks. The improvements are not limited to a single model or attack type, highlighting {\methodname}’s generalizability and adaptability to diverse adversarial scenarios.

\subsection{Cost and Privacy Comparison}\label{cost-comparison}
While GPT-4 indeed leads in many scenarios, the performance gap between EP-Shield and GPT-4 is minimal, despite GPT-4's significantly larger parameter size (EP-Shield has only 7B parameters). The GPT-4-1106-preview model used in this paper costs \textbf{\$0.03/1k tokens}. While the cost of a single query with GPT-4 may not seem high, adversarial attacks rarely involve a single query. Taking ALERT's attack on CodeBERT in the Clone Detection task as an example, the \textbf{average number of model queries exceeds 2,000 per example}, with an \textbf{average token length of about 300 tokens}~\cite{beam}. Under these conditions, using GPT-4 would cost approximately \$18 per adversarial example generation, whereas EP-Shield demonstrates far greater cost efficiency. Additionally, EP-Shield is \textbf{more suitable for privacy-sensitive scenarios} compared to commercial models like GPT-4.

\section{Conclusion}
This work establishes code naturalness as a measurable defense criterion against adversarial attacks on CLMs. Through systematic analysis of $36$ attack scenarios, we demonstrate that even SOTA identifier replacement attacks consistently generate detectable unnatural examples (NES $\le 2/5$). These results confirm that naturalness-aware defenses enhance adversarial robustness while preserving code quality, making them a viable approach for strengthening CLMs against real-world attacks with varying behaviors and vulnerabilities.

\section*{Acknowledgements}

We are deeply grateful to Yao Wan, Dongping Chen, and Hongyu Zhang for their generous assistance, insightful advice, and continuous encouragement throughout the development of this work.

\clearpage
\clearpage

%
%
\newpage
\bibliographystyle{splncs04}
\bibliography{reference}

\end{sloppypar}
\end{document}

%% file: tab/evaluation_score.tex
\begin{table*}[!t]
\centering
\caption{Proportion(\%) of NES for four attack methods on different models and tasks combinations. Avg. refers to the weighted average score of NES (i.e., the sum of scores multiplied by their respective proportions).}
\setlength{\tabcolsep}{2pt} 

\resizebox{\textwidth}{!}{
\begin{tabular}{|c|c|c c c c c c|c c c c c c|} 
\hline
\multirow{2}{*}{\textbf{Models}} & \multirow{2}{*}{\textbf{Tasks}} & \multicolumn{6}{c|}{\textbf{ALERT}} & \multicolumn{6}{c|}{\textbf{Beam-Attack}} \\ 
\cline{3-14}
 & & 1 & 2 & 3 & 4 & 5 & \textbf{Avg.} & 1 & 2 & 3 & 4 & 5 & \textbf{Avg.} \\ 
\hline
\multirow{3}{*}{CodeBERT} & Clone Detection & \textbf{45.36}  & \textbf{46.68}  & 5.44  & 2.25  & 0.27  & 1.65 & \textbf{32.06}  & \textbf{53.82}  & 10.74  & 2.79  & 0.59  & 1.86 \\
 & Code Summarization & \textbf{31.13}  & \textbf{44.96}  & 14.34  & 8.19  & 1.38  & 2.04 & \textbf{19.28}  & \textbf{38.02}  & 24.36  & 14.96  & 3.37  & 2.45 \\
 & Vulnerability Detection & \textbf{66.22}  & \textbf{29.73}  & 2.25  & 1.80  & 0.00  & 1.40 & \textbf{25.00}  & \textbf{60.94}  & 14.06  & 0.00  & 0.00  & 1.89 \\ 
\hline
\multirow{3}{*}{CodeGPT} & Clone Detection & \textbf{41.82 } & \textbf{49.41}  & 7.46  & 0.95  & 0.36  & 1.69 & \textbf{27.06}  & \textbf{59.26}  & 10.15  & 3.53  & 0.00  & 1.90 \\
 & Code Summarization & \textbf{29.56}  & \textbf{45.13}  & 15.97  & 7.42  & 1.92  & 2.07 & \textbf{18.60}  & \textbf{38.25}  & 23.34  & 14.92  & 4.89  & 2.49 \\
 & Vulnerability Detection & \textbf{38.60}  & \textbf{47.72}  & 10.03  & 3.65  & 0.00  & 1.79 & \textbf{30.33}  & \textbf{59.84}  & 9.02  & 0.82  & 0.00  & 1.80 \\ 
 \hline
\multirow{3}{*}{PLBART} & Clone Detection & \textbf{43.58}  & \textbf{47.74}  & 7.23  & 1.08  & 0.36  & 1.67 & \textbf{30.70}  & \textbf{55.70}  & 11.91  & 1.51  & 0.17  & 1.85 \\
 & Code Summarization & \textbf{27.15}  & \textbf{42.90}  & 17.60  & 10.27  & 2.08  & 2.17 & \textbf{15.94}  & \textbf{37.11}  & 21.92  & 19.00  & 6.03  & 2.62 \\
 & Vulnerability Detection & \textbf{44.23}  & \textbf{45.40}  & 7.72  & 1.48  & 1.16  & 1.70 & \textbf{23.06}  & \textbf{53.47}  & 13.06  & 5.10  & 5.31  & 2.16 \\ 
\hline
\multicolumn{2}{|c|}{Overall} & & & & & & 1.94 & & & & & & 2.33 \\ 
\hline 
 & & \multicolumn{6}{c|}{\textbf{MHM}} & \multicolumn{6}{c|}{\textbf{WIR}} \\ 
\cline{1-14}
\multirow{3}{*}{CodeBERT} & Clone Detection & \textbf{87.09}  & \textbf{12.49}  & 0.35  & 0.07  & 0.00  & 1.13 & \textbf{81.36}  & \textbf{17.75}  & 0.80  & 0.09  & 0.00  & 1.20 \\
 & Code Summarization & \textbf{79.70}  & \textbf{18.95}  & 1.18  & 0.18  & 0.00  & 1.22 & \textbf{77.85}  & \textbf{20.41}  & 1.47  & 0.26  & 0.00  & 1.24 \\
 & Vulnerability Detection & \textbf{97.10}  & \textbf{2.83}  & 0.00  & 0.00  & 0.00  & 1.03 & \textbf{89.90}  & \textbf{8.92}  & 1.01  & 0.17  & 0.00  & 1.11 \\ 
\hline
\multirow{3}{*}{CodeGPT} & Clone Detection & \textbf{84.88}  & \textbf{14.60}  & 0.52  & 0.00  & 0.00  & 1.16 & \textbf{85.21}  & \textbf{14.25}  & 0.47  & 0.07  & 0.00  & 1.15 \\
 & Code Summarization & \textbf{83.41}  & \textbf{15.04}  & 1.29  & 0.17  & 0.09  & 1.18 & \textbf{79.25}  & \textbf{18.83}  & 1.52  & 0.35  & 0.04  & 1.23 \\
 & Vulnerability Detection & \textbf{97.02}  & \textbf{2.31}  & 0.52  & 0.15  & 0.00  & 1.04 & \textbf{87.46}  & \textbf{11.16}  & 1.07  & 0.31  & 0.00  & 1.14 \\ 
\hline
\multirow{3}{*}{PLBART} & Clone Detection & \textbf{87.73}  & \textbf{11.83}  & 0.44  & 0.00  & 0.00  & 1.13 & \textbf{81.78}  & \textbf{17.29}  & 0.70  & 0.23  & 0.00  & 1.19 \\
 & Code Summarization & \textbf{74.71}  & \textbf{22.76}  & 2.21  & 0.24  & 0.09  & 1.28 & \textbf{74.22}  & \textbf{22.87}  & 2.41  & 0.46  & 0.04  & 1.29 \\
 & Vulnerability Detection & \textbf{95.79}  & \textbf{3.90}  & 0.24  & 0.00  & 0.08  & 1.05 & \textbf{93.81}  & \textbf{5.84}  & 0.35  & 0.00  & 0.00  & 1.07 \\ 
\hline 
\multicolumn{2}{|c|}{Overall} & & & & & & 1.16 & & & & & & 1.18 \\ 
\hline 
\end{tabular}
}

\label{tab:eval_score}
\end{table*}

%% file: tab/statisitcs_of_datasets.tex
\begin{table}[!t]
\centering
\caption{Statistics of the datasets.Here, CD, VD, and CS refer to Clone Detection, Vulnerability Detection, and Code Summarization, respectively, which are three types of SE tasks. P, R, F1, and acc represent Precision, Recall, F1-Score, and Accuracy, respectively.}
\begin{tabular}{|c|c|c|c|c|c|}
    \hline
    Task & Dataset Name    & Train Size & Dev Size & Test Size & Metrics   \\ 
    \hline
    CD   & BigCloneBench   & 90,102     & 4,000    & 4,000     & P, R, F1  \\
    VD   & OWASP Benchmark & 13,041     & 4,000    & 4,000     & Acc       \\
    CS   & CodeSearchNet   & 164,923    & 5,183    & 10,955    & Bleu-4    \\
    \hline
\end{tabular}

\label{tab:statisitcs of datasets}
\end{table}

%% file: tab/performance_of_CLMs.tex
\begin{table}[t]
\centering
\caption{Performance of CLMs on downstream tasks.}
\begin{tabular}{|c|c|c|c|c|c|} 
\hline
Tasks    & \multicolumn{3}{c|}{Clone Detection} & Vulnerability Detection & Code Summarization  \\ 
\hline
Models   & P      & R      & F1                 & Acc                     & BLEU-4              \\ 
\hline
CodeBERT & 0.9731 & 0.9728 & 0.9727             & 0.9890                  & 18.72               \\
CodeGPT  & 0.9688 & 0.9677 & 0.9677             & 0.9895                  & 14.78               \\
PLBART   & 0.9631 & 0.9627 & 0.9627             & 0.9962                  & 17.27               \\
\hline
\end{tabular}

\label{tab:performance of CLMs}
\end{table}

%% file: tab/attack_performance.tex
\begin{table}[!t]
\centering
\caption{Attack success rate of identifier substitution attacks across different task-victim model combinations.}
\begin{tabular}{|c|c|c|c|c|c|} 
    \hline
    \multirow{2}{*}{Tasks} & \multirow{2}{*}{Victim Models} & \multicolumn{4}{c|}{Attack Methods}         \\ 
    \cline{3-6} 
                           &                                & ALERT   & Beam-Attack & MHM     & WIR      \\ 
    \hline
    \multirow{3}{*}{Clone Detection}    & CodeBERT                       & 11.07\% & 17.52\%     & 36.59\% & 28.70\%  \\
                           & CodeGPT                        & 13.02\% & 17.49\%     & 48.89\% & 37.91\%  \\
                           & PLBART                         & 10.45\% & 15.47\%     & 46.11\% & 34.67\%  \\ 
    \hline
    \multirow{3}{*}{Vulnerability Detection}    & CodeBERT                       & 1.67\%  & 1.62\%      & 41.94\% & 15.01\%  \\
                           & CodeGPT                        & 5.05\%  & 6.15\%      & 33.90\% & 16.52\%  \\
                           & PLBART                         & 15.21\% & 12.27\%     & 31.57\% & 21.51\%  \\ 
    \hline
    \multirow{3}{*}{Code Summarization}    & CodeBERT                       & 39.43\% & 44.69\%     & 90.55\% & 70.54\%  \\
                           & CodeGPT                        & 31.28\% & 32.91\%     & 92.25\% & 60.83\%  \\
                           & PLBART                         & 59.43\% & 65.34\%     & 91.80\% & 64.86\%  \\
    \hline
\end{tabular}

\label{tab: ASR of methods}
\end{table}

%% file: tab/consistency_analysis.tex
\begin{table}[!t]
\centering
\caption{Consistency analysis of NES across {\methodmetric}, GPT-4, and human evaluations at various analytical perspectives.
Each cell contains three values: the exact match rate ($1$st), the consistency rate within a tolerance of 1 ($2$nd), and the mean absolute difference ($3$rd).
}
\small
\begin{tabular}{|c|c|c|c|} 
\hline
\multicolumn{4}{|c|}{\textit{Task Level}}                                                                                            \\ 
\hline
Tasks                                                                              &       & GPT-4                 & Human                  \\ 
\hline
\multirow{2}{*}{\begin{tabular}[c]{@{}l@{}}Clone Detection\end{tabular}}         & GPT-4 & -                     & 0.77 / 0.99  \\
& {\methodmetric}  & 0.75 / 0.99  & 0.75 / 0.99  \\ 
\hline
\multirow{2}{*}{\begin{tabular}[c]{@{}l@{}}Vulnerability Detection\end{tabular}} & GPT-4 & -                     & 0.87 / 1.00   \\
& {\methodmetric}  & 0.85 / 0.99  & 0.84 / 0.99   \\ 
\hline
\multirow{2}{*}{\begin{tabular}[c]{@{}l@{}}Code Summarization\end{tabular}}      & GPT-4 & -                     & 0.63 / 0.94  \\
& {\methodmetric}  & 0.60 / 0.92 & 0.67 / 0.96  \\ 
\hline \hline
\multicolumn{4}{|c|}{\textit{Victim Model Level}}                                                                                    \\ 
\hline
Models                                                                              &       & GPT-4                 & Human                  \\ 
\hline
\multirow{2}{*}{CodeBERT}                                                          & GPT-4 & -                     & 0.73 / 0.99 \\
& {\methodmetric}  & 0.69 / 0.97  & 0.69 / 0.97   \\ 
\hline
\multirow{2}{*}{CodeGPT}                                                           & GPT-4 & -                     & 0.67 / 0.96  \\
& {\methodmetric}  & 0.63 / 0.93  & 0.71 / 0.96   \\ 
\hline
\multirow{2}{*}{PLBART}                                                            & GPT-4 & -                     & 0.69 / 0.98   \\
& {\methodmetric}  & 0.64 / 0.96  & 0.64 / 0.96   \\ 
\hline \hline
\multicolumn{4}{|c|}{\textit{Attack Method Level}}                                                                                    \\ 
\hline
Methods                                                                            &       & GPT-4                 & Human                  \\ 
\hline
\multirow{2}{*}{ALERT}                                                             & GPT-4 & -                     & 0.49 / 0.90  \\
& {\methodmetric}  & 0.45 / 0.90 & 0.55 / 0.94  \\ 
\hline
\multirow{2}{*}{Beam-Attack}                                                       & GPT-4 & -                     & 0.57 / 0.97  \\
& {\methodmetric}  & 0.51 / 0.93 & 0.50 / 0.93   \\ 
\hline
\multirow{2}{*}{MHM}                                                               & GPT-4 & -                     & 0.86 / 1.00  \\
& {\methodmetric}  & 0.85 / 1.00  & 0.86 / 0.99  \\ 
\hline
\multirow{2}{*}{WIR}                                                               & GPT-4 & -                     & 0.82 / 1.00   \\
& {\methodmetric}  & 0.81 / 0.99  & 0.80 / 0.99   \\ 
\hline
\end{tabular}

\label{tab:consistency-analysis}
\vspace{-1em}
\end{table}

%% file: tab/correlation_overall.tex
\begin{table}[!t]
\centering
\caption{Correlations between {\methodmetric} and other adversarial example quality metrics and human performance, with $r_{s}$, $r_{p}$, and $\tau$ representing Spearman, Pearson, and Kendall-Tau correlation coefficients. 
}
\setlength{\tabcolsep}{4pt}
\setlength{\extrarowheight}{0pt}
\addtolength{\extrarowheight}{\aboverulesep}
\addtolength{\extrarowheight}{\belowrulesep}
\setlength{\aboverulesep}{0pt}
\setlength{\belowrulesep}{0pt}
\begin{tabular}{|c|c c c|c|} 
\hline
Metrics                    & $r_s$                                      & $r_p$                                      & $\tau$                                     & Avg.                                        \\ 
\hline
ICR ($\downarrow$)         & -0.1860                                    & -0.2748                                    & -0.1683                                    & -0.2097                                     \\
TCR ($\downarrow$)         & -0.3140                                    & -0.3187                                    & -0.2664                                    & -0.2997                                     \\
ACS ($\uparrow$)           & {0.4639}                                     & \textcolor{gray}{0.4674}                                     & 0.3770                                     & 0.4361                                      \\
AED ($\downarrow$)         & -0.4660                                    & -0.4498                                    & \textcolor{gray}{-0.3938}                                    & \textcolor{gray}{-0.4365}                                    \\
PPL ($\downarrow$)       & -0.0637                                    & 0.0290                                     & -0.0491                                    & -0.0279                                     \\
CodeBLEU ($\uparrow$)      & 0.0014                                     & 0.1104                                     & 0.0123                                     & 0.0414                                      \\
CodeBERTScore ($\uparrow$) & \textcolor{gray}{0.4757}                                     & 0.4445                                     & 0.3893                                     & \textcolor{gray}{0.4365}                                      \\ 
\midrule
{\methodmetric} ($\uparrow$)               & \textbf{0.6687} & \textbf{0.7419} & \textbf{0.6386} & \textbf{0.6831}  \\
\bottomrule
\end{tabular}

\label{tab:correlation metric vs human}
\vspace{-1em}
\end{table}

%% file: tab/purify_compare_subset.tex
\begin{table}[!t]
\centering
\caption{Comparison of the defense success rates of adversarial fine-tuning (Adv-tuning), GPT-$4$ and {\methodpurifier} on four attack methods against the CodeBERT model across three tasks. $\Delta$ represents the percentage improvement of {\method} compared to Adv-tuning. 
}
\setlength{\tabcolsep}{3pt}
\begin{tabular}{|l|cccc|c|} 
\hline
\multicolumn{1}{|l|}{\multirow{2}{*}{Defence}} & \multicolumn{4}{c|}{Attack Methods}        & \multirow{2}{*}{Overall}  \\ 
\cline{2-5}
\multicolumn{1}{|l|}{}                         & ALERT   & Beam-Attack & MHM     & WIR     &                           \\ 
\hline
\multicolumn{6}{|c|}{\textit{Task 1:~ Clone Detection}}                                                                \\ 
\hline
Adv-tuning                                   & 89.71   & 90.16       & 87.90   & 89.52   & 89.32                     \\
GPT-4                                        & \textcolor{gray}{95.76}   & \textcolor{gray}{96.49}       & \textcolor{gray}{97.50}   & \textcolor{gray}{97.15}   & \textcolor{gray}{96.73}                     \\
{\methodpurifier}                                         & \textbf{98.89}   & \textbf{98.48}       & \textbf{98.58}   & \textbf{99.27}   & \textbf{98.81}                     \\
$\Delta$                        & \textcolor{red}{+ 9.18}  & \textcolor{red}{+ 8.32}      & \textcolor{red}{+ 10.68} & \textcolor{red}{+ 9.75}  & \textcolor{red}{+ 9.48}                    \\ 
\hline
\multicolumn{6}{|c|}{\textit{Task 2:~ Vulnerability Detection}}                                                        \\ 
\hline
Adv-tuning                                   & 92.40   & 89.16       & 91.02   & 96.30   & 92.22                     \\
GPT-4                                        & \textcolor{gray}{98.65}   & \textcolor{gray}{98.44}       & \textcolor{gray}{97.35}   & \textcolor{gray}{98.48}   & \textcolor{gray}{98.23}                     \\
{\methodpurifier}                                         & \textbf{97.27}   & \textbf{93.75}       & \textbf{95.10}   & \textbf{97.61}   & \textbf{95.93}                     \\
$\Delta$                      & \textcolor{red}{+ 4.87}  & \textcolor{red}{+ 4.59}      & \textcolor{red}{+ 4.08}  & \textcolor{red}{+ 1.31}  & \textcolor{red}{+ 3.71 }                   \\ 
\hline
\multicolumn{6}{|c|}{\textit{Task 3:~ Code Summarization}}                                                             \\ 
\hline
Adv-tuning                                   & 80.78   & 76.58       & 73.88   & 67.70   & 74.74                     \\
GPT-4                                        & \textcolor{gray}{91.60}   & \textcolor{gray}{89.89}       & \textcolor{gray}{94.71}   & \textcolor{gray}{91.86}   & \textcolor{gray}{92.02}                     \\
{\methodpurifier}                                         & \textbf{91.28}   & \textbf{87.89}       & \textbf{93.11}   & \textbf{92.05}   & \textbf{91.08}                     \\
$\Delta$                        & \textcolor{red}{+ 10.50} & \textcolor{red}{+ 11.31}     & \textcolor{red}{+ 19.23} & \textcolor{red}{+ 24.35} & \textcolor{red}{+ 16.35}                   \\
\hline
\end{tabular}

\label{table:defense compare on codebert}
\vspace{-1em}
\end{table}

%% file: tab/misclassfication_rate.tex
\begin{table}[!t]
\centering
\caption{{\methodmetric} defines the misclassification rates (\%) of original samples as those under attack.
}
\begin{tabular}{|c|c c c c|} 
\hline
\multicolumn{1}{|c|}{Tasks}  & $\delta = 4$ & $\delta = 3$ & $\delta = 2$ & $\delta = 1$ \\ 
\hline
Clone Detection                   & 64.30 & 15.44 & 6.07 & 2.44 \\
Code Summarization                  & 60.52 & 16.76 & 9.47 & 2.23 \\
Vulnerability Detection                   & 62.27 & 7.29 & 3.91 & 0.58 \\
\hline
\end{tabular}

\label{tab: misclassfication-rate}
\vspace{-1em}
\end{table}

%% file: tab/defence4unknown.tex
\begin{table}[!t]
\caption{Comparison of the defense success rate (\%) of {\methodpurifier} and Adv-tuning against RNNS on three different models in the vulnerability detection task. 
}
\centering
\small
\begin{tabular}{|c|cc|c|} 
\hline
Victim Models & Adv-tuning & \methodpurifier  & $\Delta$  \\ 
\hline
CodeBert      & 73.02      & 99.50 & \textcolor{red}{+26.48}                  \\
CodeGPT       & 66.00      & 97.34 & \textcolor{red}{+31.34}                  \\
PLBART        & 82.35      & 87.32 & \textcolor{red}{+4.97}                   \\
\hline
\end{tabular}

\label{tab:RNNS-VD}
\vspace{-1em}
\end{table}

%% file: tab/defence4unknown_2.tex
\begin{table}[!t]
\centering
\caption{Comparison of defense success rates (\%) of {\methodpurifier} and Adv-tuning against 3 attack methods on CodeBERT in the Authorship Attribution task. 
}
\begin{tabular}{|c|cc|c|} 
\hline
Attack Methods & Adv-tuning & {\methodpurifier}  & $\Delta$  \\ 
\hline
MHM            & 58.43      & 70.76 & \textcolor{red}{+12.33}  \\
ALERT          & 65.38      & 76.92 & \textcolor{red}{+11.54}  \\
RNNS           & 60.92      & 68.97 & \textcolor{red}{+8.05}   \\
\hline
\end{tabular}

\label{tab:RNNS-AS}
\vspace{-1em}
\end{table}

%% file: tab/purify_compare.tex
\begin{table*}[!t]

\centering

\caption{The comparison of defense success rate between {\method}, GPT-4, and adversarial fine-tuning (referred to as Adv-tuning). Downward arrow(${\downarrow}$) and upward arrow(${\uparrow}$) represent performance differences compared to {\method}. Adv-tuning refers to adversarial tuning.}

\resizebox{\linewidth}{!}{
\begin{tabular}{|c|c|cccc|c|} 
\hline
\multirow{2}{*}{Victim  Model} & \multicolumn{1}{c|}{\multirow{2}{*}{Defence}} & \multicolumn{4}{c|}{Attack Method} & \multirow{2}{*}{Overall}  \\ 
\cline{3-6}
 & \multicolumn{1}{c|}{} & ALERT & Beam-Attack & MHM & WIR &  \\ 
\hline
\multicolumn{7}{|l|}{\textit{Task 1:~ Clone-Detection }} \\ 
\hline
\multirow{3}{*}{{CodeBERT}} 
& Adv-tuning & 89.71\%~$_{\downarrow9.18\%}$ & 90.16\%~$_{\downarrow8.32\%}$ & 87.90\%~$_{\downarrow10.68\%}$ & 89.52\%~$_{\downarrow9.75\%}$ & 89.32\%$_{\downarrow9.48\%}$ \\ 
& GPT-4 & 95.76\%~$_{\downarrow3.13\%}$ & 96.49\%~$_{\downarrow1.99\%}$ & 97.50\%~$_{\downarrow1.08\%}$ & 97.15\%~$_{\downarrow2.12\%}$ & 96.73\% $_{\downarrow2.08\%}$ \\
& {\method} & 98.89\% & 98.48\% & 98.58\% & 99.27\% & 98.81\% \\
\hline
\multirow{3}{*}{{CodeGPT}}  
& Adv-tuning & 15.64\%~$_{\downarrow83.63\%}$ & 20.59\%~$_{\downarrow74.79\%}$ & 23.98\%~$_{\downarrow75.07\%}$ & 23.07\%~$_{\downarrow76.10\%}$ & 20.82\%~$_{\downarrow77.40\%}$ \\ 
& GPT-4 & 97.63\%~$_{\downarrow1.64\%}$ & 95.44\%~$_{\uparrow0.06\%}$ & 98.59\%~$_{\downarrow0.46\%}$ & 98.64\%~$_{\downarrow0.53\%}$ & 97.58\%~$_{\downarrow0.64\%}$ \\
& {\method} & 99.27\% & 95.38\% & 99.05\% & 99.17\% & 98.22\% \\
\hline
\multirow{3}{*}{{PLBART}}   
& Adv-tuning & 88.12\%~$_{\downarrow10.42\%}$ & 88.52\%~$_{\downarrow8.74\%}$ & 90.89\%~$_{\downarrow8.33\%}$ & 91.55\%~$_{\downarrow7.98\%}$ & 89.77\%$_{\downarrow8.87\%}$ \\ 
& GPT-4 & 98.92\%~$_{\uparrow0.38\%}$ & 97.82\%~$_{\uparrow 0.56\%}$ & 99.44\%~$_{\uparrow 0.22\%}$ & 99.88\%~ $_{\uparrow 0.35\%}$ & 99.02\%~$_{\uparrow 0.38\%}$ \\
& {\method} & 98.54\% & 97.26\% & 99.22\% & 99.53\% & 98.64\% \\
\hline
\multicolumn{7}{|l|}{\textit{Task 2:~ Vulnerability-Detection }} \\ 
\hline
\multirow{3}{*}{{CodeBERT}} 
& Adv-tuning & 92.40\%~$_{\downarrow4.87\%}$ & 89.16\%~$_{\downarrow4.59\%}$ & 91.02\%~$_{\downarrow4.08\%}$ & 96.30\%~$_{\downarrow1.31\%}$ & 92.22\%~$_{\downarrow3.71\%}$ \\ 
& GPT-4 & 98.65\%~$_{\uparrow 1.38\%}$ & 98.44\%~$_{\uparrow 4.69\%}$ & 97.35\%~$_{\uparrow 2.25\%}$ & 98.48\%~$_{\uparrow 0.87\%}$ & 98.23\%$_{\uparrow 2.30\%}$ \\
& {\method} & 97.27\% & 93.75\% & 95.10\% & 97.61\% & 95.93\% \\
\hline
\multirow{3}{*}{{CodeGPT}}  
& Adv-tuning & 87.64\%~$_{\downarrow5.99\%}$ & 87.03\%~$_{\downarrow4.27\%}$ & 88.87\%~$_{\downarrow6.06\%}$ & 90.27\%~$_{\downarrow3.42\%}$ & 88.45\%~$_{\downarrow4.94\%}$ \\ 
& GPT-4 & 95.44\%~$_{\uparrow 1.81\%}$ & 93.85\%~$_{\uparrow 2.55\%}$ & 96.88\%~$_{\uparrow 1.95\%}$ & 96.94\%~$_{\uparrow 3.25\%}$ & 95.78\%~$_{\uparrow 2.39\%}$ \\
& {\method} & 93.63\% & 91.30\% & 94.93\% & 93.69\% & 93.39\% \\
\hline
\multirow{3}{*}{{PLBART}}   
& Adv-tuning & 87.62\%~$_{\uparrow 0.54\%}$ & 86.03\%~$_{\uparrow 1.90\%}$ & 89.53\%~$_{\downarrow0.29\%}$ & 86.85\%~$_{\downarrow1.09\%}$ & 87.51\%~$_{\uparrow 0.27\%}$ \\ 
& GPT-4 & 81.80\%~$_{\downarrow5.28\%}$ & 84.90\%~$_{\uparrow 0.77\%}$ & 90.62\%~$_{\uparrow 0.80\%}$ & 86.93\%~$_{\downarrow1.01\%}$ & 86.06\%~$_{\downarrow1.18\%}$ \\
& {\method} & 87.08\% & 84.13\% & 89.82\% & 87.94\% & 87.24\% \\
\hline
\multicolumn{7}{|l|}{\textit{Task 3:~ Code-Summarization }} \\ 
\hline
\multirow{3}{*}{{CodeBERT}} 
& Adv-tuning & 80.78\%~$_{\downarrow10.50\%}$ & 76.58\%~$_{\downarrow11.31\%}$ & 73.88\%~$_{\downarrow19.23\%}$ & 67.70\%~$_{\downarrow24.35\%}$ & 74.74\%~$_{\downarrow16.35\%}$ \\ 
& GPT-4 & 91.60\%~$_{\uparrow 0.32\%}$ & 89.89\%~$_{\uparrow 2.00\%}$ & 94.71\%~$_{\uparrow 1.60\%}$ & 91.86\%~$_{\downarrow0.19\%}$ & 92.02\%~$_{\uparrow 0.93\%}$ \\
& {\method} & 91.28\% & 87.89\% & 93.11\% & 92.05\% & 91.08\% \\
\hline
\multirow{3}{*}{{CodeGPT}}  
& Adv-tuning & 74.96\%~$_{\downarrow12.29\%}$ & 72.27\%~$_{\downarrow11.65\%}$ & 58.85\%~$_{\downarrow36.19\%}$ & 57.36\%~$_{\downarrow32.41\%}$ & 65.86\%~$_{\downarrow23.14\%}$ \\ 
& GPT-4 & 87.34\%~$_{\uparrow 0.09\%}$ & 86.76\%~$_{\uparrow 2.84\%}$ & 93.93\%~$_{\downarrow1.11\%}$ & 90.99\%~$_{\uparrow 1.22\%}$ & 89.76\%~$_{\uparrow 0.76\%}$ \\
& {\method} & 87.25\% & 83.92\% & 95.04\% & 89.77\% & 89.00\% \\
\hline
\multirow{3}{*}{{PLBART}}   
& Adv-tuning & 82.18\%~$_{\downarrow7.51\%}$ & 83.76\%~$_{\downarrow5.36\%}$ & 79.15\%~$_{\downarrow12.79\%}$ & 75.77\%~$_{\downarrow12.29\%}$ & 80.22\%~$_{\downarrow9.49\%}$ \\
& GPT-4 & 90.86\%~$_{\uparrow 1.17\%}$ & 90.79\%~$_{\uparrow 1.67\%}$ & 92.54\%~$_{\uparrow 0.60\%}$ & 89.83\%~$_{\uparrow1.77\%}$ & 91.01\%~$_{\uparrow 1.30\%}$ \\
& {\method} & 89.69\% & 89.12\% & 91.94\% & 88.06\% & 89.70\% \\
\hline
\end{tabular}
}

\label{table:defense compare}
\end{table*}

%% file: tab/preft_and_postft.tex
\begin{table}[!t]
\centering
\caption{Comparison of defense success rates of {\method} before and after Fine-Tuning. Pre-FT and Post-FT denote performance before and after fine-tuning, respectively. Upward arrow ($\uparrow$) indicates the extent of improvement post-FT.}

\begin{tabular}{|c|c|c|c|} 
    \hline
    Victim Models     & \multicolumn{1}{c|}{Attack Methods} & Pre-FT & Post-FT \\ 
    \hline
    \multirow{4}{*}{CodeBERT} & ALERT                                      & 80.51\%         & 98.89\%~$_{\uparrow18.38\%}$           \\
                              & Beam-Attack                                & 77.73\%         & 98.48\%~$_{\uparrow20.75\%}$            \\
                              & MHM                                        & 67.47\%         & 98.58\%~$_{\uparrow34.11\%}$            \\
                              & WIR                                        & 73.35\%         & 99.27\%~$_{\uparrow25.92\%}$            \\ 
    \hline
    \multirow{4}{*}{CodeGPT}  & ALERT                                      & 79.83\%         & 99.27\%~$_{\uparrow19.44\%}$           \\
                              & Beam-Attack                                & 76.34\%         & 95.38\%~$_{\uparrow19.04\%}$            \\
                              & MHM                                        & 70.79\%         & 99.05\%~$_{\uparrow28.26\%}$            \\
                              & WIR                                        & 73.67\%         & 99.17\%~$_{\uparrow25.50\%}$            \\ 
    \hline
    \multirow{4}{*}{PLBART}   & ALERT                                      & 78.06\%         & 98.54\%~$_{\uparrow20.48\%}$           \\
                              & Beam-Attack                                & 76.95\%         & 97.26\%~$_{\uparrow20.31\%}$            \\
                              & MHM                                        & 68.17\%         & 99.22\%~$_{\uparrow31.05\%}$            \\
                              & WIR                                        & 72.61\%         & 99.53\%~$_{\uparrow26.92\%}$            \\
    \hline
    \end{tabular}

\label{tab: pre-ft-vs.-post-ft}
\end{table}